\documentclass[prb,superscriptaddress,amsmath,amssymb, notitlepage, twocolumn,nofootinbib,longbibliography]{revtex4-1}

\pdfoutput=1

\usepackage{amsmath}
\usepackage{tabularx,graphicx}
\usepackage{epstopdf}
\usepackage{graphicx}
\usepackage{latexsym}
\usepackage{amssymb}
\usepackage{amsmath}
\usepackage{color, colortbl}
\usepackage{psfrag}
\usepackage{bbm}
\usepackage{bm}
\usepackage{titlesec}
\usepackage{dsfont}
\usepackage{feynmp}
\usepackage{slashed}
\usepackage{multirow}

\usepackage[papersize={8.5in,11in}]{geometry}

\definecolor{darkblue}{rgb}{0.,0.,0.4}
\definecolor{darkred}{rgb}{0.5,0.,0.}
\definecolor{BlueViolet}{RGB}{138,43,226}
\definecolor{SkyBlue}{RGB}{30,144,255}
\definecolor{DarkGreen}{RGB}{0,100,0}
\usepackage[pdftex,colorlinks=true,linkcolor=darkblue,citecolor=blue,urlcolor=darkred]{hyperref}

\usepackage{url}
\usepackage{color}
\usepackage{gensymb}
\DeclareMathOperator\erfc{erfc}
\usepackage[tight]{subfigure}

\begin{document}
\title{Correlated Insulators in Twisted Bilayer Graphene}
\author{Ipsita Mandal}
\affiliation{Faculty of Science and Technology, University of Stavanger, 4036 Stavanger, Norway}
\affiliation{Nordita, Roslagstullsbacken 23, SE-106 91 Stockholm, Sweden}
\affiliation{Laboratory of Atomic And Solid State Physics, Cornell University, Ithaca, NY 14853, USA}
\author{Jia Yao}
\affiliation{
Department of Physics,
California Institute of Technology, Pasadena, CA 91125, USA
}
\author{ Erich J. Mueller}
\affiliation{Laboratory of Atomic And Solid State Physics, Cornell University, Ithaca, NY 14853, USA}
\date{\today}

\begin{abstract}
Experiments on graphene bilayers, where the top layer is rotated with respect to the one below, have displayed insulating behavior when the moir\'e bands are partially filled.  We calculate the charge distributions in these phases, and estimate the excitation gaps.
\end{abstract}

\maketitle

\section{Introduction}

Graphene bilayers, where the top layer is rotated with respect to the bottom, show remarkable properties \cite{cao1,cao2,choi,zondiner,sharpe,jiang,kerelsky,yazdani,kimdasilva,yankowitz}. These arise from the presence of a long wavelength moir\'e superlattice.  For twists near certain ``magic angles", the low energy bands become very flat, and interactions dominate\cite{Bistritzer}.  Moreover, the bands have non-trivial topological indices.  Experimentally one observes insulating phases at certain rational fillings of the bands.  Electrostatically doping away from these rational fillings leads to superconducting phases, whose transition temperatures are large compared to the bandwidth \cite{cao1,cao2}.  An important part of understanding the physics of these systems is to identify the structure of the correlated insulating states.  Here we conduct a variational study of the various possible charge-density, spin-density, and valley-density waves, which are the most natural candidates.  We find that there are ``stripe" ordered spin and valley ferromagnets at fillings of $\lbrace 1/8, 3/8, 5/8, 7/8 \rbrace$.  At fillings of $\lbrace 1/4,1/2,3/4 \rbrace$, the ground state is a spatially homogeneous spin and valley ferromagnet.  The stripe order should give rise to birefringence.

There is significant prior work on this problem.  Two separate groups, Seo, Kotov, and Uchoa \cite{seo}, and, Kang and Vafek \cite{kang} recently argued for a ferromagnetic state at $1/8$ filling.  Kang and Vafek's results are similar to ours, in that they find charge ordering in addition to the spin ordering.  Our approach is complementary in that we explore different models for the electron-electron interactions.  Kang and Vafek use a short-range interaction, while we consider two models: a long-range Coulomb interaction, and a dipolar interaction which accounts for image charges in the back-gate.  All three models give similar results, pointing to the robustness of the phenomena.

Related analysis is found in the supplementary material of Choi et al. \cite{choi}, which considers the half-filled case. Lu et al. \cite{efetov} found signatures of an orbital ferromagnetic state at $3/8$ filling.
Studies of Wigner crystallization by Padhi, Setty, and Phillips \cite{philip1,philip2} have relevance.
Also notable is the slave-spin treatment of Pizarro, Calderon, and Bascones \cite{pizarro}, the Hartree-Fock analysis of Xie and MacDonald \cite{xiemacdonald}, and the projection-based technique of Repellin, Dong, Zhang, and Senthil \cite{repelin}.  All of these reveal different aspects of the correlated insulators.

In addition to those already described \cite{cao1,cao2}, experimental studies of the correlated insulators include measurements of the compressibility \cite{choi,zondiner}, magnetic response \cite{sharpe}, tunneling spectroscopy \cite{jiang,kerelsky,yazdani} and further transport properties \cite{kimdasilva,yankowitz}.

Beyond finding the lowest energy charge configurations, we investigate the energy cost of adding a particle, adding a hole, or adding a particle-hole pair.  We find that the interaction energy from these defects is large compared to the width of the moir\'e mini-bands.  Consequently, one expects large interaction-driven mixing in of higher bands, which would need to be included in quantitative models.  Our present study, however, is a necessary prerequisite for those calculations.

\begin{figure}[tb]
\includegraphics[width=0.9\columnwidth]{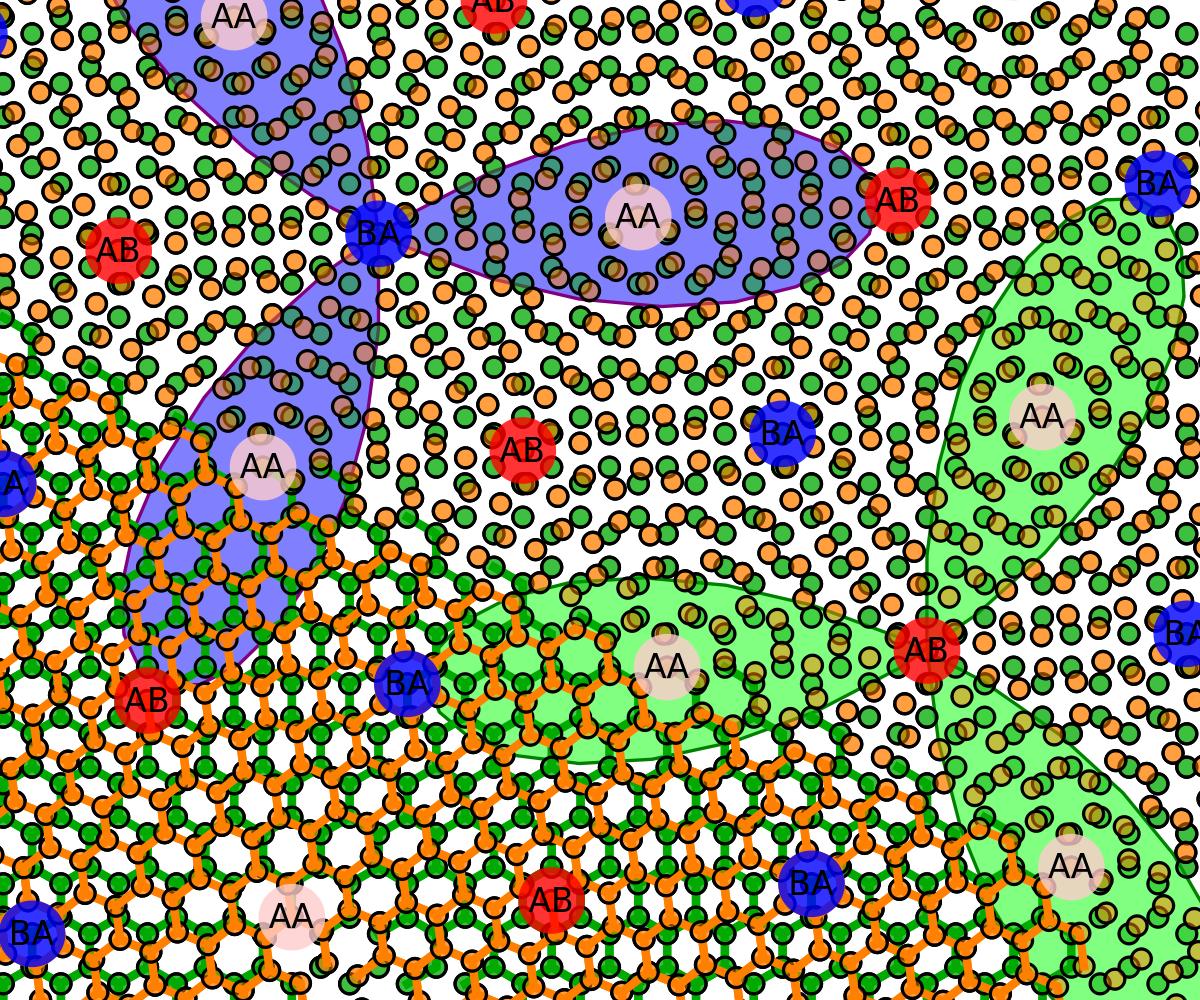}
\caption{(Color Online)
Schematic of twisted bilayer graphene: Green/orange discs represent carbon atoms in each layer.  In the lower left, lines connect nearest-neighbor atoms in each layer.  $AA$ (pink), $AB$ (red), and $BA$ (blue) regions are labeled.  Two schematic trefoil-shaped Wannier states are shown, which centered on $BA$ and $AB$ regions respectively.  
}\label{geo}
\end{figure}

\section{Model}
Graphene forms a honeycomb lattice, with a unit cell containing two sites -- denoted $A$ and $B$.  For small twist-angles (near $1$\degree), the bilayer system displays a large moir\'e unit cell.  There are three notable regions, denoted by $AA$, $AB$, and $BA$, which each forming triangular lattices (see Fig.~\ref{geo}).  In the $AA$ regions, the two lattices align.  In the $AB$ regions, the $A$ sites of the lower layer line up with the $B$ sites of the upper layer.  The $BA$ region is the opposite. The $AB$ and $BA$ sites together form a honeycomb lattice.

As described in Ref.~\onlinecite{koshino-fu}, the electronic structure of the low energy bands is built from Wannier states which are centered at the $AB$ and $BA$ sites. Each Wannier state has a three-lobed spatial structure, with each lobe centered on an $AA$ region.  Thus the Wannier state centers are in the $AB$-$BA$ regions, but the charge density is peaked in the $AA$ regions.  This structure is schematically shown in Fig.~\ref{geo}.
In addition to a sublattice index ($AB$ or $BA$), the Wannier states in this model are labeled by spin and valley indices, leading to an eight-band effective Hamiltonian. 

There is a number of subtleties about the construction of the Wannier states.  For example, 
Refs.~\onlinecite{zhouvish,zou2,vafek-wang} argued that there is a topological obstruction that prevents the existence of exponentially localized Wannier states obeying all the emergent symmetries. Indeed, while the Wannier states from Ref.~\onlinecite{koshino-fu} that we work with are exponentially localized, one of the protecting symmetries is broken \cite{zou2,vafek-wang}. These considerations only affect physics on an exponentially small energy scale, where symmetry breaking effects matter: They are not relevant to the questions we are asking, and our results would be unchanged if we worked with a larger basis set.

The effective electronic hopping matrix element between two nearest Wannier orbitals is about $0.3$ meV, while the effective direct Coulomb interaction between them is of the order of $ 10$ meV \cite{koshino-fu,seo}.  This clear separation of scales motivates us to consider stationary charge distributions, where we break translational symmetry, and distribute the electrons among various Wannier orbitals.  We neglect kinetic energy and find the configurations which minimize the Coulomb interaction energy.

We systematically consider periodic arrangements of charges defined by the generators of the supercell ${\bf v}_1=m_1\, {\bf a}_1+m_2\, {\bf a}_2 $ and ${\bf v}_2=n_1\, {\bf a}_1+n_2 \,{\bf a}_2$,
where ${\bf a}_1$ and ${\bf a}_2$ are the generators of the moir\'e lattice.  We include all unit cells containing fewer than 13 sublattice sites, with each of $m_1,m_2,n_1,$ and $n_2$ restricted to be smaller than or equal to 4.  We also include a small number of larger unit cells. In all, we consider
well over 3000 charge configurations -- though many of them are equivalent, or symmetry related.

We use the interaction model derived  by Koshino et al. \cite{koshino-fu}.  They found that  the Coulomb interaction energy from lobes in the $AA$ regions $\bf r_i$ and $\bf r_j$  is well approximated by
\begin{align}
   V_{ij} =   \begin{cases}
   \frac{e^2}{9\,\epsilon \,L_M} 
   \frac{1}{|r_i-r_j|}&r_{i}\neq r_j\\
   \frac{e^2}{9\times 0.28 \,L_M} & r_{i}=r_j
   \end{cases} \,,
\end{align}
where $-e/3$ is the charge in each lobe, $\epsilon$ is the dielectric constant, which we take equal to $10\,\epsilon_0$.
Here $L_M$ is the distance between neighboring $AB$ and $BA$ regions.
We use $L_M= 13.4$ nm, corresponding to a twist angle of  $\theta = 1.05$\degree.
We also include an exchange interaction, which lowers the energy by $J$ when two overlapping occupied orbitals have the same spin and valley quantum  numbers.  We use the values for the exchange energies from Ref.~\onlinecite{koshino-fu}.  These are dominated by the nearest neighbor and next-nearest neighbor terms: $J=0.376, 0.0645\,e^2/(\epsilon\,L_M)$. All other exchange energies are smaller than $0.0014\,e^2/(\epsilon\,L_M)$. Terms beyond the fifth-nearest neighbour are ignored.

The Coulomb interaction energy from a two-dimensional (2D) array of charges is linearly divergent with system size. The relative energies of the configurations with the same average density, however, are well defined.  Some care must be taken in calculating these energies, since even after subtracting off the leading divergence, the remaining sums are conditionally convergent.  The process is regularized by using 2D Ewald  sums \cite{ewald,ewaldlee}:
\begin{align}
\label{ewc}
&\sum_{{\mathbf{w}}\in {\cal L}} \frac{1}{|{\mathbf{r}} - { \mathbf{w}}|}
= f_s({\bf  r})+f_l({\bf  r})+C,\\
&f_l = \frac{2\pi}{\Omega} 
\sum_{{\bf  k}\in \bar{\cal L}\backslash0} 
\frac{e^{ \mathrm{i} \,{\bf {k}\cdot 
{r}}}}{2\,k} 
\erfc \left (k \, \eta \right ), \label{fl}
\\\label{fs}
&f_s = \sum_{{\bf  w}\in {\cal L}}\frac{1}{|{\bf  r -  w}|}\erfc\left(\frac{|{\bf  r -  w}|}{2\eta}\right)
- \frac{4\sqrt{\pi}}{\Omega}\eta.
\end{align}
Here $\eta$ is an arbitrary parameter which separates the interactions into long-range and short-range parts. The sum is over a Bravais lattice $\cal L$, which in our case is generated by the supecell vectors $\mathbf{ v_1,v_2}$.  
The reciprocal lattice is $\bar {\cal L}$, and  $\bar{\cal L}\backslash0$ represents the reciprocal lattice with the origin removed.  Here, $C$ is an irrelevant infinite constant, which physically represents the Coulomb energy from a uniform charge distribution.  The sums in Eqs.~(\ref{fl}) and (\ref{fs}) are absolutely convergent.
We chose $\eta$ to give a reasonably fast convergence rate, typically taking it to be $1/3$ of the lattice constant.

In the experimental set-up of
Ref.~\onlinecite{cao2}, 
a metallic back-gate sits roughly $30$ nm behind the sample.  We supplement our long-range Coulomb calculations by also calculating energies where we include a set of image charges in this layer.  The Appendix gives the resulting expressions for the energies.  To leading order, these images uniformly shift the energies of all configurations by the same amount (corresponding to the classical capacitance).  Configuration-dependent corrections are exponentially small in the ratio of the supercell periodicity to the distance from the back-gate, and we find that for the experimental geometry the results for ground state energies are unchanged when we add the images.

\begin{figure}[t]
\includegraphics[width = 0.35\textwidth]{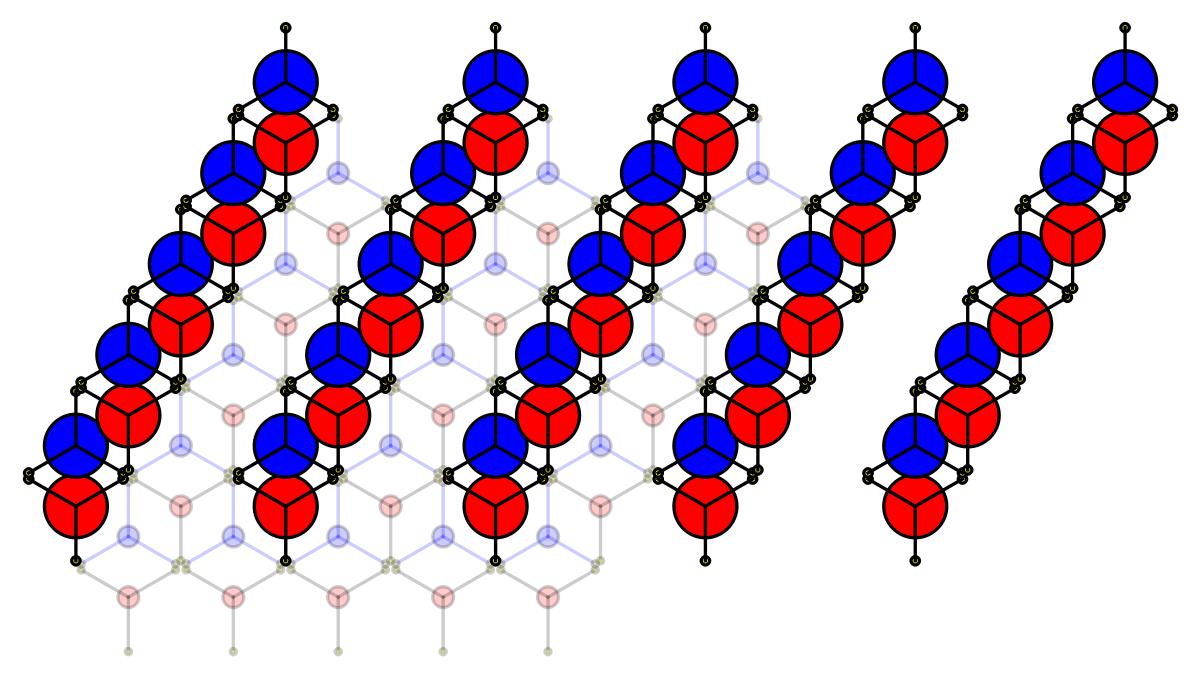}
\caption{(color online) The ground state for $1/8$ filling. 
The large red and blue discs represent electrons occupying the Wannier states on the $AB$ and $BA$ lattice sites.  Each Wannier state has a three-lobed structure -- and lines extend from each colored disc -- terminated by small black dots which represent the center of each lobe.  Small pale discs show the locations of empty Wannier states for part of the lattice.
Despite a breaking of translational invariance,  the charge density is uniform:  each $AA$ region has a charge of $-e$.
}
\label{fig:gs}
\end{figure}

\section{Ground States}

As previously explained, there are eight states per unit cell -- labeled by sublattice ($AB$ or $BA$), spin, and valley.
Due to the relatively strong exchange interaction, physically corresponding to large spatial overlap between neighboring Wannier states, it is always favorable to maximize the number of electrons which have the same spin and valley quantum numbers.  Thus we find that at fillings $\nu=1/8,1/4$, all electrons have the same spin and valley quantum numbers.  At $\nu=3/8$, there are not enough states for all electrons to carry the same spin and valley quantum numbers.  Instead, two-thirds of the electrons carry one set, while the other one-third carry another set. At $\nu=1/2$, things are similar, but an equal number of electrons carry each of the two sets of quantum numbers.  Larger $\nu$'s are simply particle-hole mirrors of these configurations.
At the level of our model, the exact quantum numbers do not matter -- what matters is just that they follow this pattern.

We find that for $\nu=1/8$, the occupied modes (all of which have the same spin and valley quantum numbers) form a striped configuration, as shown in Fig.~\ref{fig:gs}. 
For $\nu=1/4$, the occupations are uniform, with every Wannier state occupied by a single electron -- all of which share the same spin and valley quantum numbers. 
For $\nu=3/8$, the electrons with the dominant quantum numbers form a uniform pattern, while the ones with the other quantum number form the  pattern from $\nu=1/8$.  For $\nu=1/2$, the occupations are again uniform, with each Wannier site being doubly occupied.

\begin{figure}[t]
\includegraphics[width = 0.42\textwidth]{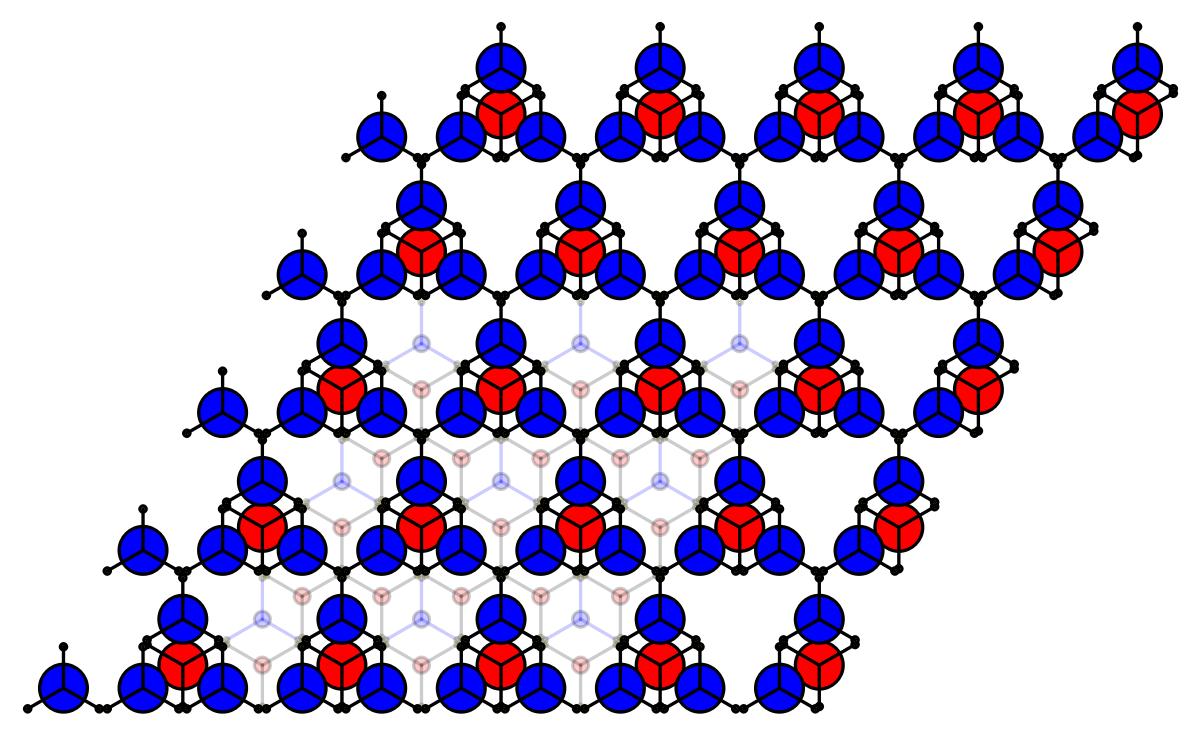}
\caption{(color online) Second-lowest energy periodic configuration that we found with a  filling of $1/8$.    If one allows larger unit cells, one can make lower energy excited states by alternating regions of this pattern with those from Fig.~\ref{fig:gs}.}
\label{ring}
\end{figure}

 Our results are confirmed by the experimental observations of  Zondiner et al. \cite{zondiner}, where they find that right at an integer filling, the spin and valley degrees of freedom are not filled equally. Starting with $\nu=1/8$, only one flavour fills it up. Then at the next integer filling, another additional flavor fills it up, and so on.

Despite the symmetry breaking at $\nu=1/8$, the charge density is nearly uniform:  Each $AA$ site has a charge of $-e$.  The spatial symmetry breaking, however, should be apparent in optical measurements through a birefringence.  One may also find that transport measurements are anisotropic.

The energy per particle in each of these four cases is
$E =-1.72,-2.39,35.5,49.4 \, {e^2}/
 \left( \epsilon\, L_M \right).$
In physical units, this corresponds to
$E=-18.4,-25.6,381,530$ meV per particle. 
The relatively small difference between the energy per particle for $\nu=1/8$ and $\nu=1/4$ occurs because in these spin/valley polarized states, the direct Coulomb interaction is largely cancelled by the exchange interactions.
At higher fillings, we are adding electrons with different spin/valley quantum numbers,
for which there is no such cancellation.  These energies should all be understood as relative to a classical uniform charge distribution.

One way to quantify the stability of the striped state at $\nu=1/8$, is to look at the next-lowest energy state that we found (Fig.~\ref{ring}). This excited configuration has energy
$-17.7$ meV per particle, which is only modestly higher than the ground state.  The charge density for this pattern is again uniform, even though the pattern of occupied Wannier states is non-uniform. Note, one can create $\nu=1/8$ configurations whose energies lie between these by alternating these two patterns. 

Given the small energy difference, lattice defects, impurities, or other irregularities could play a role in the experimentally observed pattern.  Furthermore, kinetic effects could mix them.

\section{Excitations}

\label{secexite}

We calculate the energy of four types of excitations: adding a single electron to one of our ground states, removing a single electron (i.e., adding a hole), adding an electron-hole pair, or flipping the spin/valley of a single electron: $E_{\rm e},E_{\rm h},E_{\rm eh},E_{\rm s}$.  Physically, $E_{\rm e}$ and $E_{\rm h}$  correspond to the chemical potentials required to add or remove a charge and are experimentally determined by measuring the charge density as a function of gate voltage:  the charge density is constant over a voltage range of width $\Delta V=(E_{\rm e}+E_{\rm h})/e$.   The electron-hole excitation energy can experimentally be measured from transport:  one expects the low temperature resistivity to obey an Arrhenius law with energy scale $E_{\rm eh}$.  The spin excitation energy can similarly be inferred from the temperature dependence of the spin susceptibility.
While we find that the presence of a back-gate plays no role in the ground state energies, it can shift the excitations energies by $\sim 1$ meV.  We just quote numbers for the case without the back-gate.

\begin{figure}[t]
\includegraphics[width = 0.35 \textwidth]{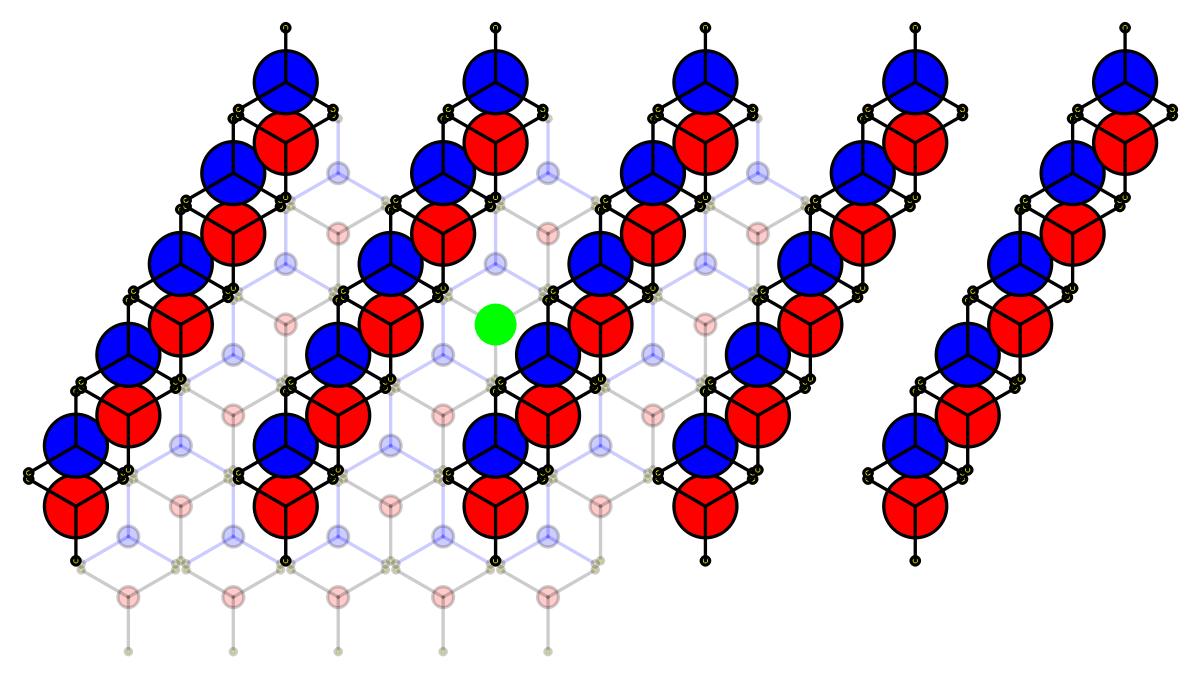}
\caption{The addition of an electron to an empty site in a ground state configuration for $1/8$ filling. The green dot represents the added electron.}
\label{fig:excite}
\end{figure}

Experiments find the energy scales for excitations are fractions of meV, while our model yields tens of meV.  The discrepancy has a number of sources.  First, since the Coulomb interaction scale is large compared to the width of the moir\'e mini-bands, the low energy excitations undoubtedly involve electronic states which are outside of the minibands (and hence not included in our model).  Second, it is likely that impurities and other defects are playing a role in the experiment.  Third, collective effects can lower the excitation energy.  Nonetheless, our estimates are quite valuable as a starting point for understanding the phenomena.

For $\nu=1/8$, the lowest energy electron excitation consists of adding a particle with the same quantum numbers as the other atoms to one of the empty sites (as shown in Fig.~\ref{fig:excite}): all possibilities are equivalent.  This has energy 
$E_{\rm e}=-1.33
\times \frac{e^2}{\epsilon\,L_M}=-14.3$ meV. Similarly, all ways of removing an electron are equivalent, with $E_{\rm h}=36.9$ meV.
As is required for stability, $-E_{\rm h}<E_{\rm e}$, which is a discrete analogy to requiring that the compressibility is positive, $\partial^2 E/\partial N^2>0$.
The fact that $E_{\rm e}<0$ should not cause alarm --  this is a common physical occurrence.  The physical picture is that our energies are measured relative to having a uniform charge distribution, and we effectively have contributions from a uniform positive background.

The lowest energy particle-hole excitation at $\nu=1/8$ occurs when the particle is moved to a neighboring unoccupied site.  This has the energy $E_{\rm eh}=10.2$ meV, which is smaller than $E_{\rm e}+E_{\rm h}$ because of the electron-hole interactions.  It is also larger than the energy-per-particle difference between the configurations in Fig.~\ref{fig:gs} and Fig.~\ref{ring}.  One can generate Fig.~\ref{ring} from Fig.~\ref{fig:gs} by moving one-fourth of the particles.  The interactions between these particle-hole pairs reduce the cost of this rearrangement.
A spin or valley flip excitation costs an energy of $E_{\rm s}= 10.1$ meV, which is solely due to the exchange interaction.  

For $\nu=1/4$, any extra electron must have different quantum numbers than the others.  All sites are equivalent, and we find $E_e=-13.8 $ meV.
A naive expectation might be that this excitation energy would be twice what we found at $\nu=1/8$, since the charge is twice as large.  In fact adding an electron at $\nu=1/4$ costs much more than twice the $\nu=1/8$ value.  Given that the charge density is very same in both cases, this difference can be attributed to the exchange interaction.  The energy to remove an electron is $E_{\rm h}= 51.2 $ meV, which is somewhat less than twice the energy of the equivalent excitation at $\nu=1/8$.  The difference can be attributed to the detailed structure of the charge distribution in the Wannier states.  The particle-hole excitation energy is $E_{\rm eh} = 21.1$ meV.
A spin or valley flip excitation costs an energy of 
$E_s = 17.5$ meV.

At $\nu=3/8$, the ground state contains electrons with two different quantum  numbers. Let us denote these by $\alpha$ and $\beta$.  The $\alpha$ electrons are spread uniformly, while the $\beta$ electrons form the zigzag pattern in Fig.~\ref{fig:gs}. The lowest energy electronic excitation corresponds to adding another electron in the $\beta$ state, with an  energy of $E_{ \rm e}=-28.1$ meV.
Clearly, this is just  the sum of the values of $E_{\rm e}$ for the $\nu=1/8$ and $\nu=1/4$ configurations.
Note that if we add an electron with a different spin/valley, the value would have been $-20.7$ meV, which is energetically costlier. The lowest energy hole excitation comes from removing an electron from the $\beta$ state, which yields $E_{\rm h}= 33.2$ meV. The particle-hole excitation energy is $E_{\rm eh} = 10.2$ meV. A spin or valley flip excitation costs an energy of $E_{ \rm s} =10.1$ meV.

At $\nu=1/2$, any added electron must have different quantum numbers than the rest of electrons. The required energy is $E_{\rm e}= -27.6$ meV. As expected, this comes out to be twice the value of $E_{\rm e}$ for $\nu=1/4$.
The energy required to remove an electron is $E_{\rm h}=65.0$ meV.
The particle-hole excitation energy is $E_{\rm eh} =21.1$ meV.
A spin or valley flip excitation costs an energy of $E_{\rm s} = 17.5$ meV. We note that the value of $E_{\rm s}$ is the same for $\nu=1/8$ and $\nu=3/8$ (equal to $10.1$ meV), and for $\nu=1/4$ and $\nu=1/2$ (equal to $17.5$ meV).

Much of the structure in the excitation energies is related to the exchange interaction: The direct Coulomb energy (in units of $\frac{e^2}{\epsilon\,L_M}$) for adding a particle to the grounds states corresponding to $ 1/8,1/4,3/8,1/2$ fillings are: $0.642, 1.284, 1.926, 2.568$, which are in the ratio $1:2:3:4$. This scaling arises because the charge distribution is uniform in all cases.

\section{Summary and Outlook}

To summarize, we have calculated the charge distributions of the correlated insulators at all integer fillings of magic angle twisted bilayer graphene.
We have also estimated all possible excitation gaps over the insulating ground states. In all these computations, we have neglected the kinetic terms, as the effective tight-binding models show that the strength of the
hopping terms is much much less than the Coulomb interactions, and will only have a perturbative effect. 
One essential caveat is that because the Coulomb scale is so large, it likely leads to mixing between the flat bands and the spectator bands.  Nonetheless we are able to explain much of the observed physics, such as the progression of fillings seen by Zondiner et al.\cite{zondiner}

It would be rewarding to apply similar techniques 
to characterize the correlated insulating phases of transition metal dichalcogenide (TMD) homobilayers and heterobilayers \cite{PhysRevLett.121.026402,PhysRevLett.122.086402,2019arXiv191008673T,2020Natur.579..359R,fu-tmd}, as they have
much simpler moir\'e band structures. Moreover, in certain TMDs such as WSe$_2$, the flat bands and the resulting  correlated  states are found to exist over a continuum of angles \cite{cory-wse2} rather than a narrow range around some ``magic angles".

\section{Acknowledgements}
This work was supported by the NSF Grant PHY-1806357.


\appendix

\section*{Interactions with Image Charges}
\label{appdipole}

We model screening by a back-gate sitting a distance $d/2$ from the sample by placing equal and opposite image charges at a distance $d$.  We treat the charge distribution of each Wannier state as three delta functions, each of charge $q=e/3$.
The Coulomb interaction between a single charge in one plane, and a Bravais lattice of charges in another is 
\begin{equation}
    V=\lambda \sum_{{\bf w}\in {\cal L}} \frac{1}{\sqrt{|{\bf r-w}|^2+d^2}}\,,
\end{equation}
where $\lambda=q^2/(4\pi\epsilon )$,
${\bf w}$ is a 2D lattice vector, and $\bf r$ is the 2D location of the charge in the device.  By Fourier transforming the potential, one can write:
\begin{equation}
   V= \lambda \,C+\frac{2\pi \lambda \,d}{\Omega}+
   \frac{2\pi \lambda}{\Omega}
   \sum_{{\bf q}\in \bar{\cal L}\backslash0}\frac{e^{-q\, d}}{|q|} e^{\mathrm{i}\,{\bf q\cdot r}}\, ,
\end{equation}
where the first term contains the same infinite constant as Eq.~(\ref{ewc}).  The second term is the potential from a uniformly charged plane, with  $\Omega$ being the area of the unit cell.  The third term falls off exponentially with the separation. There, $\bar{\cal L}$ is the 2D reciprocal lattice.  Due to this exponential cutoff, the image charges play a very minor role in the energetics of different periodic charge configurations.

We follow the procedure in Lee and Cai \cite{ewaldlee} to write $V=V_s+V_l$ with
\begin{widetext}
\begin{align}
V_s =&
\sum_{w\in{\cal L}}
\frac{ \erfc \left (
\frac{\sqrt{|r-w|^2+d^2}} { 2\,\eta} \right )}
{\sqrt{|r-w|^2+d^2}}
+
\frac{  2\pi d\,
\erfc \left (\frac{d}{2\,\eta}\right ) 
- 4\,\sqrt{\pi}\, \eta \,e^{-\frac{ d^2} {4\,\eta^2}}}
{\Omega}
\,,
\nonumber \\
V_l =&
\frac{2\,\sqrt{\pi}}{\Omega}
\sum_{{\bf \vec k}\in \bar{\cal L}\backslash0} 
\left[
2 e^{-k\delta} 
-e^{-k\delta}\erfc\left(\frac{\delta}{2\eta}-k\eta \right )
+ e^{ k\delta}
\erfc\left(\frac{\delta}{2\eta}+k\eta\right)
\right] .
\end{align}
\end{widetext}

\bibliography{biblio}

\begin{thebibliography}{31}%
\makeatletter
\providecommand \@ifxundefined [1]{%
 \@ifx{#1\undefined}
}%
\providecommand \@ifnum [1]{%
 \ifnum #1\expandafter \@firstoftwo
 \else \expandafter \@secondoftwo
 \fi
}%
\providecommand \@ifx [1]{%
 \ifx #1\expandafter \@firstoftwo
 \else \expandafter \@secondoftwo
 \fi
}%
\providecommand \natexlab [1]{#1}%
\providecommand \enquote  [1]{``#1''}%
\providecommand \bibnamefont  [1]{#1}%
\providecommand \bibfnamefont [1]{#1}%
\providecommand \citenamefont [1]{#1}%
\providecommand \href@noop [0]{\@secondoftwo}%
\providecommand \href [0]{\begingroup \@sanitize@url \@href}%
\providecommand \@href[1]{\@@startlink{#1}\@@href}%
\providecommand \@@href[1]{\endgroup#1\@@endlink}%
\providecommand \@sanitize@url [0]{\catcode `\\12\catcode `\$12\catcode
  `\&12\catcode `\#12\catcode `\^12\catcode `\_12\catcode `\%12\relax}%
\providecommand \@@startlink[1]{}%
\providecommand \@@endlink[0]{}%
\providecommand \url  [0]{\begingroup\@sanitize@url \@url }%
\providecommand \@url [1]{\endgroup\@href {#1}{\urlprefix }}%
\providecommand \urlprefix  [0]{URL }%
\providecommand \Eprint [0]{\href }%
\providecommand \doibase [0]{http://dx.doi.org/}%
\providecommand \selectlanguage [0]{\@gobble}%
\providecommand \bibinfo  [0]{\@secondoftwo}%
\providecommand \bibfield  [0]{\@secondoftwo}%
\providecommand \translation [1]{[#1]}%
\providecommand \BibitemOpen [0]{}%
\providecommand \bibitemStop [0]{}%
\providecommand \bibitemNoStop [0]{.\EOS\space}%
\providecommand \EOS [0]{\spacefactor3000\relax}%
\providecommand \BibitemShut  [1]{\csname bibitem#1\endcsname}%
\let\auto@bib@innerbib\@empty
\bibitem [{\citenamefont {{Cao}}\ \emph
  {et~al.}(2018{\natexlab{a}})\citenamefont {{Cao}}, \citenamefont {{Fatemi}},
  \citenamefont {{Fang}}, \citenamefont {{Tomarken}}, \citenamefont {{Luo}},
  \citenamefont {{Sanchez-Yamagishi}}, \citenamefont {{Watanabe}},
  \citenamefont {{Taniguchi}}, \citenamefont {{Kaxiras}}, \citenamefont
  {{Ashoori}},\ and\ \citenamefont {{Jarillo-Herrero}}}]{cao1}%
  \BibitemOpen
  \bibfield  {author} {\bibinfo {author} {\bibfnamefont {Y.}~\bibnamefont
  {{Cao}}}, \bibinfo {author} {\bibfnamefont {V.}~\bibnamefont {{Fatemi}}},
  \bibinfo {author} {\bibfnamefont {S.}~\bibnamefont {{Fang}}}, \bibinfo
  {author} {\bibfnamefont {S.~L.}\ \bibnamefont {{Tomarken}}}, \bibinfo
  {author} {\bibfnamefont {J.~Y.}\ \bibnamefont {{Luo}}}, \bibinfo {author}
  {\bibfnamefont {J.~D.}\ \bibnamefont {{Sanchez-Yamagishi}}}, \bibinfo
  {author} {\bibfnamefont {K.}~\bibnamefont {{Watanabe}}}, \bibinfo {author}
  {\bibfnamefont {T.}~\bibnamefont {{Taniguchi}}}, \bibinfo {author}
  {\bibfnamefont {E.}~\bibnamefont {{Kaxiras}}}, \bibinfo {author}
  {\bibfnamefont {R.~C.}\ \bibnamefont {{Ashoori}}}, \ and\ \bibinfo {author}
  {\bibfnamefont {P.}~\bibnamefont {{Jarillo-Herrero}}},\ }\bibfield  {title}
  {\enquote {\bibinfo {title} {{Correlated insulator behaviour at half-filling
  in magic-angle graphene superlattices}},}\ }\href {\doibase
  10.1038/nature26154} {\bibfield  {journal} {\bibinfo  {journal} {Nature}\
  }\textbf {\bibinfo {volume} {556}},\ \bibinfo {pages} {80--84} (\bibinfo
  {year} {2018}{\natexlab{a}})}\BibitemShut {NoStop}%
\bibitem [{\citenamefont {{Cao}}\ \emph
  {et~al.}(2018{\natexlab{b}})\citenamefont {{Cao}}, \citenamefont {{Fatemi}},
  \citenamefont {{Fang}}, \citenamefont {{Watanabe}}, \citenamefont
  {{Taniguchi}}, \citenamefont {{Kaxiras}},\ and\ \citenamefont
  {{Jarillo-Herrero}}}]{cao2}%
  \BibitemOpen
  \bibfield  {author} {\bibinfo {author} {\bibfnamefont {Y.}~\bibnamefont
  {{Cao}}}, \bibinfo {author} {\bibfnamefont {V.}~\bibnamefont {{Fatemi}}},
  \bibinfo {author} {\bibfnamefont {S.}~\bibnamefont {{Fang}}}, \bibinfo
  {author} {\bibfnamefont {K.}~\bibnamefont {{Watanabe}}}, \bibinfo {author}
  {\bibfnamefont {T.}~\bibnamefont {{Taniguchi}}}, \bibinfo {author}
  {\bibfnamefont {E.}~\bibnamefont {{Kaxiras}}}, \ and\ \bibinfo {author}
  {\bibfnamefont {P.}~\bibnamefont {{Jarillo-Herrero}}},\ }\bibfield  {title}
  {\enquote {\bibinfo {title} {{Unconventional superconductivity in magic-angle
  graphene superlattices}},}\ }\href {\doibase doi:10.1038/nature26160}
  {\bibfield  {journal} {\bibinfo  {journal} {Nature}\ }\textbf {\bibinfo
  {volume} {556}},\ \bibinfo {pages} {43--50} (\bibinfo {year}
  {2018}{\natexlab{b}})}\BibitemShut {NoStop}%
\bibitem [{\citenamefont {Choi}\ \emph {et~al.}(2019)\citenamefont {Choi},
  \citenamefont {Kemmer}, \citenamefont {Peng}, \citenamefont {Thomson},
  \citenamefont {Arora}, \citenamefont {Polski}, \citenamefont {Zhang},
  \citenamefont {Ren}, \citenamefont {Alicea}, \citenamefont {Refael},
  \citenamefont {von Oppen}, \citenamefont {Watanabe}, \citenamefont
  {Taniguchi},\ and\ \citenamefont {Nadj-Perge}}]{choi}%
  \BibitemOpen
  \bibfield  {author} {\bibinfo {author} {\bibfnamefont {Youngjoon}\
  \bibnamefont {Choi}}, \bibinfo {author} {\bibfnamefont {Jeannette}\
  \bibnamefont {Kemmer}}, \bibinfo {author} {\bibfnamefont {Yang}\ \bibnamefont
  {Peng}}, \bibinfo {author} {\bibfnamefont {Alex}\ \bibnamefont {Thomson}},
  \bibinfo {author} {\bibfnamefont {Harpreet}\ \bibnamefont {Arora}}, \bibinfo
  {author} {\bibfnamefont {Robert}\ \bibnamefont {Polski}}, \bibinfo {author}
  {\bibfnamefont {Yiran}\ \bibnamefont {Zhang}}, \bibinfo {author}
  {\bibfnamefont {Hechen}\ \bibnamefont {Ren}}, \bibinfo {author}
  {\bibfnamefont {Jason}\ \bibnamefont {Alicea}}, \bibinfo {author}
  {\bibfnamefont {Gil}\ \bibnamefont {Refael}}, \bibinfo {author}
  {\bibfnamefont {Felix}\ \bibnamefont {von Oppen}}, \bibinfo {author}
  {\bibfnamefont {Kenji}\ \bibnamefont {Watanabe}}, \bibinfo {author}
  {\bibfnamefont {Takashi}\ \bibnamefont {Taniguchi}}, \ and\ \bibinfo {author}
  {\bibfnamefont {Stevan}\ \bibnamefont {Nadj-Perge}},\ }\bibfield  {title}
  {\enquote {\bibinfo {title} {Electronic correlations in twisted bilayer
  graphene near the magic angle},}\ }\href {\doibase 10.1038/s41567-019-0606-5}
  {\bibfield  {journal} {\bibinfo  {journal} {Nature Physics}\ }\textbf
  {\bibinfo {volume} {15}},\ \bibinfo {pages} {1174--1180} (\bibinfo {year}
  {2019})}\BibitemShut {NoStop}%
\bibitem [{\citenamefont {Zondiner}\ \emph {et~al.}(2020)\citenamefont
  {Zondiner}, \citenamefont {Rozen}, \citenamefont {Rodan-Legrain},
  \citenamefont {Cao}, \citenamefont {Queiroz}, \citenamefont {Taniguchi},
  \citenamefont {Watanabe}, \citenamefont {Oreg}, \citenamefont {von Oppen},
  \citenamefont {Stern}, \citenamefont {Berg}, \citenamefont
  {Jarillo-Herrero},\ and\ \citenamefont {Ilani}}]{zondiner}%
  \BibitemOpen
  \bibfield  {author} {\bibinfo {author} {\bibfnamefont {U.}~\bibnamefont
  {Zondiner}}, \bibinfo {author} {\bibfnamefont {A.}~\bibnamefont {Rozen}},
  \bibinfo {author} {\bibfnamefont {D.}~\bibnamefont {Rodan-Legrain}}, \bibinfo
  {author} {\bibfnamefont {Y.}~\bibnamefont {Cao}}, \bibinfo {author}
  {\bibfnamefont {R.}~\bibnamefont {Queiroz}}, \bibinfo {author} {\bibfnamefont
  {T.}~\bibnamefont {Taniguchi}}, \bibinfo {author} {\bibfnamefont
  {K.}~\bibnamefont {Watanabe}}, \bibinfo {author} {\bibfnamefont
  {Y.}~\bibnamefont {Oreg}}, \bibinfo {author} {\bibfnamefont {F.}~\bibnamefont
  {von Oppen}}, \bibinfo {author} {\bibfnamefont {Ady}\ \bibnamefont {Stern}},
  \bibinfo {author} {\bibfnamefont {E.}~\bibnamefont {Berg}}, \bibinfo {author}
  {\bibfnamefont {P.}~\bibnamefont {Jarillo-Herrero}}, \ and\ \bibinfo {author}
  {\bibfnamefont {S.}~\bibnamefont {Ilani}},\ }\bibfield  {title} {\enquote
  {\bibinfo {title} {Cascade of phase transitions and dirac revivals in
  magic-angle graphene},}\ }\href {\doibase 10.1038/s41586-020-2373-y}
  {\bibfield  {journal} {\bibinfo  {journal} {Nature}\ }\textbf {\bibinfo
  {volume} {582}},\ \bibinfo {pages} {203--208} (\bibinfo {year}
  {2020})}\BibitemShut {NoStop}%
\bibitem [{\citenamefont {Sharpe}\ \emph {et~al.}(2019)\citenamefont {Sharpe},
  \citenamefont {Fox}, \citenamefont {Barnard}, \citenamefont {Finney},
  \citenamefont {Watanabe}, \citenamefont {Taniguchi}, \citenamefont
  {Kastner},\ and\ \citenamefont {Goldhaber-Gordon}}]{sharpe}%
  \BibitemOpen
  \bibfield  {author} {\bibinfo {author} {\bibfnamefont {Aaron~L.}\
  \bibnamefont {Sharpe}}, \bibinfo {author} {\bibfnamefont {Eli~J.}\
  \bibnamefont {Fox}}, \bibinfo {author} {\bibfnamefont {Arthur~W.}\
  \bibnamefont {Barnard}}, \bibinfo {author} {\bibfnamefont {Joe}\ \bibnamefont
  {Finney}}, \bibinfo {author} {\bibfnamefont {Kenji}\ \bibnamefont
  {Watanabe}}, \bibinfo {author} {\bibfnamefont {Takashi}\ \bibnamefont
  {Taniguchi}}, \bibinfo {author} {\bibfnamefont {M.~A.}\ \bibnamefont
  {Kastner}}, \ and\ \bibinfo {author} {\bibfnamefont {David}\ \bibnamefont
  {Goldhaber-Gordon}},\ }\bibfield  {title} {\enquote {\bibinfo {title}
  {Emergent ferromagnetism near three-quarters filling in twisted bilayer
  graphene},}\ }\href {\doibase 10.1126/science.aaw3780} {\bibfield  {journal}
  {\bibinfo  {journal} {Science}\ }\textbf {\bibinfo {volume} {365}},\ \bibinfo
  {pages} {605--608} (\bibinfo {year} {2019})}\BibitemShut {NoStop}%
\bibitem [{\citenamefont {Jiang}\ \emph {et~al.}(2019)\citenamefont {Jiang},
  \citenamefont {Lai}, \citenamefont {Watanabe}, \citenamefont {Taniguchi},
  \citenamefont {Haule}, \citenamefont {Mao},\ and\ \citenamefont
  {Andrei}}]{jiang}%
  \BibitemOpen
  \bibfield  {author} {\bibinfo {author} {\bibfnamefont {Yuhang}\ \bibnamefont
  {Jiang}}, \bibinfo {author} {\bibfnamefont {Xinyuan}\ \bibnamefont {Lai}},
  \bibinfo {author} {\bibfnamefont {Kenji}\ \bibnamefont {Watanabe}}, \bibinfo
  {author} {\bibfnamefont {Takashi}\ \bibnamefont {Taniguchi}}, \bibinfo
  {author} {\bibfnamefont {Kristjan}\ \bibnamefont {Haule}}, \bibinfo {author}
  {\bibfnamefont {Jinhai}\ \bibnamefont {Mao}}, \ and\ \bibinfo {author}
  {\bibfnamefont {Eva~Y.}\ \bibnamefont {Andrei}},\ }\bibfield  {title}
  {\enquote {\bibinfo {title} {Charge order and broken rotational symmetry in
  magic-angle twisted bilayer graphene},}\ }\href {\doibase
  10.1038/s41586-019-1460-4} {\bibfield  {journal} {\bibinfo  {journal}
  {Nature}\ }\textbf {\bibinfo {volume} {573}},\ \bibinfo {pages} {91--95}
  (\bibinfo {year} {2019})}\BibitemShut {NoStop}%
\bibitem [{\citenamefont {Kerelsky}\ \emph {et~al.}(2019)\citenamefont
  {Kerelsky}, \citenamefont {McGilly}, \citenamefont {Kennes}, \citenamefont
  {Xian}, \citenamefont {Yankowitz}, \citenamefont {Chen}, \citenamefont
  {Watanabe}, \citenamefont {Taniguchi}, \citenamefont {Hone}, \citenamefont
  {Dean}, \citenamefont {Rubio},\ and\ \citenamefont {Pasupathy}}]{kerelsky}%
  \BibitemOpen
  \bibfield  {author} {\bibinfo {author} {\bibfnamefont {Alexander}\
  \bibnamefont {Kerelsky}}, \bibinfo {author} {\bibfnamefont {Leo~J.}\
  \bibnamefont {McGilly}}, \bibinfo {author} {\bibfnamefont {Dante~M.}\
  \bibnamefont {Kennes}}, \bibinfo {author} {\bibfnamefont {Lede}\ \bibnamefont
  {Xian}}, \bibinfo {author} {\bibfnamefont {Matthew}\ \bibnamefont
  {Yankowitz}}, \bibinfo {author} {\bibfnamefont {Shaowen}\ \bibnamefont
  {Chen}}, \bibinfo {author} {\bibfnamefont {K.}~\bibnamefont {Watanabe}},
  \bibinfo {author} {\bibfnamefont {T.}~\bibnamefont {Taniguchi}}, \bibinfo
  {author} {\bibfnamefont {James}\ \bibnamefont {Hone}}, \bibinfo {author}
  {\bibfnamefont {Cory}\ \bibnamefont {Dean}}, \bibinfo {author} {\bibfnamefont
  {Angel}\ \bibnamefont {Rubio}}, \ and\ \bibinfo {author} {\bibfnamefont
  {Abhay~N.}\ \bibnamefont {Pasupathy}},\ }\bibfield  {title} {\enquote
  {\bibinfo {title} {Maximized electron interactions at the magic angle in
  twisted bilayer graphene},}\ }\href {\doibase 10.1038/s41586-019-1431-9}
  {\bibfield  {journal} {\bibinfo  {journal} {Nature}\ }\textbf {\bibinfo
  {volume} {572}},\ \bibinfo {pages} {95--100} (\bibinfo {year}
  {2019})}\BibitemShut {NoStop}%
\bibitem [{\citenamefont {Xie}\ \emph {et~al.}(2019)\citenamefont {Xie},
  \citenamefont {Lian}, \citenamefont {J{\"a}ck}, \citenamefont {Liu},
  \citenamefont {Chiu}, \citenamefont {Watanabe}, \citenamefont {Taniguchi},
  \citenamefont {Bernevig},\ and\ \citenamefont {Yazdani}}]{yazdani}%
  \BibitemOpen
  \bibfield  {author} {\bibinfo {author} {\bibfnamefont {Yonglong}\
  \bibnamefont {Xie}}, \bibinfo {author} {\bibfnamefont {Biao}\ \bibnamefont
  {Lian}}, \bibinfo {author} {\bibfnamefont {Berthold}\ \bibnamefont
  {J{\"a}ck}}, \bibinfo {author} {\bibfnamefont {Xiaomeng}\ \bibnamefont
  {Liu}}, \bibinfo {author} {\bibfnamefont {Cheng-Li}\ \bibnamefont {Chiu}},
  \bibinfo {author} {\bibfnamefont {Kenji}\ \bibnamefont {Watanabe}}, \bibinfo
  {author} {\bibfnamefont {Takashi}\ \bibnamefont {Taniguchi}}, \bibinfo
  {author} {\bibfnamefont {B.~Andrei}\ \bibnamefont {Bernevig}}, \ and\
  \bibinfo {author} {\bibfnamefont {Ali}\ \bibnamefont {Yazdani}},\ }\bibfield
  {title} {\enquote {\bibinfo {title} {Spectroscopic signatures of many-body
  correlations in magic-angle twisted bilayer graphene},}\ }\href {\doibase
  10.1038/s41586-019-1422-x} {\bibfield  {journal} {\bibinfo  {journal}
  {Nature}\ }\textbf {\bibinfo {volume} {572}},\ \bibinfo {pages} {101--105}
  (\bibinfo {year} {2019})}\BibitemShut {NoStop}%
\bibitem [{\citenamefont {Kim}\ \emph {et~al.}(2017)\citenamefont {Kim},
  \citenamefont {DaSilva}, \citenamefont {Huang}, \citenamefont {Fallahazad},
  \citenamefont {Larentis}, \citenamefont {Taniguchi}, \citenamefont
  {Watanabe}, \citenamefont {LeRoy}, \citenamefont {MacDonald},\ and\
  \citenamefont {Tutuc}}]{kimdasilva}%
  \BibitemOpen
  \bibfield  {author} {\bibinfo {author} {\bibfnamefont {Kyounghwan}\
  \bibnamefont {Kim}}, \bibinfo {author} {\bibfnamefont {Ashley}\ \bibnamefont
  {DaSilva}}, \bibinfo {author} {\bibfnamefont {Shengqiang}\ \bibnamefont
  {Huang}}, \bibinfo {author} {\bibfnamefont {Babak}\ \bibnamefont
  {Fallahazad}}, \bibinfo {author} {\bibfnamefont {Stefano}\ \bibnamefont
  {Larentis}}, \bibinfo {author} {\bibfnamefont {Takashi}\ \bibnamefont
  {Taniguchi}}, \bibinfo {author} {\bibfnamefont {Kenji}\ \bibnamefont
  {Watanabe}}, \bibinfo {author} {\bibfnamefont {Brian~J.}\ \bibnamefont
  {LeRoy}}, \bibinfo {author} {\bibfnamefont {Allan~H.}\ \bibnamefont
  {MacDonald}}, \ and\ \bibinfo {author} {\bibfnamefont {Emanuel}\ \bibnamefont
  {Tutuc}},\ }\bibfield  {title} {\enquote {\bibinfo {title} {Tunable moir{\'e}
  bands and strong correlations in small-twist-angle bilayer graphene},}\
  }\href {\doibase 10.1073/pnas.1620140114} {\bibfield  {journal} {\bibinfo
  {journal} {Proceedings of the National Academy of Sciences}\ }\textbf
  {\bibinfo {volume} {114}},\ \bibinfo {pages} {3364--3369} (\bibinfo {year}
  {2017})}\BibitemShut {NoStop}%
\bibitem [{\citenamefont {Yankowitz}\ \emph {et~al.}(2019)\citenamefont
  {Yankowitz}, \citenamefont {Chen}, \citenamefont {Polshyn}, \citenamefont
  {Zhang}, \citenamefont {Watanabe}, \citenamefont {Taniguchi}, \citenamefont
  {Graf}, \citenamefont {Young},\ and\ \citenamefont {Dean}}]{yankowitz}%
  \BibitemOpen
  \bibfield  {author} {\bibinfo {author} {\bibfnamefont {Matthew}\ \bibnamefont
  {Yankowitz}}, \bibinfo {author} {\bibfnamefont {Shaowen}\ \bibnamefont
  {Chen}}, \bibinfo {author} {\bibfnamefont {Hryhoriy}\ \bibnamefont
  {Polshyn}}, \bibinfo {author} {\bibfnamefont {Yuxuan}\ \bibnamefont {Zhang}},
  \bibinfo {author} {\bibfnamefont {K.}~\bibnamefont {Watanabe}}, \bibinfo
  {author} {\bibfnamefont {T.}~\bibnamefont {Taniguchi}}, \bibinfo {author}
  {\bibfnamefont {David}\ \bibnamefont {Graf}}, \bibinfo {author}
  {\bibfnamefont {Andrea~F.}\ \bibnamefont {Young}}, \ and\ \bibinfo {author}
  {\bibfnamefont {Cory~R.}\ \bibnamefont {Dean}},\ }\bibfield  {title}
  {\enquote {\bibinfo {title} {Tuning superconductivity in twisted bilayer
  graphene},}\ }\href {\doibase 10.1126/science.aav1910} {\bibfield  {journal}
  {\bibinfo  {journal} {Science}\ }\textbf {\bibinfo {volume} {363}},\ \bibinfo
  {pages} {1059--1064} (\bibinfo {year} {2019})}\BibitemShut {NoStop}%
\bibitem [{\citenamefont {Bistritzer}\ and\ \citenamefont
  {MacDonald}(2011)}]{Bistritzer}%
  \BibitemOpen
  \bibfield  {author} {\bibinfo {author} {\bibfnamefont {Rafi}\ \bibnamefont
  {Bistritzer}}\ and\ \bibinfo {author} {\bibfnamefont {Allan~H.}\ \bibnamefont
  {MacDonald}},\ }\bibfield  {title} {\enquote {\bibinfo {title} {Moir{\'e}
  bands in twisted double-layer graphene},}\ }\href {\doibase
  10.1073/pnas.1108174108} {\bibfield  {journal} {\bibinfo  {journal}
  {Proceedings of the National Academy of Sciences}\ }\textbf {\bibinfo
  {volume} {108}},\ \bibinfo {pages} {12233--12237} (\bibinfo {year}
  {2011})}\BibitemShut {NoStop}%
\bibitem [{\citenamefont {Seo}\ \emph {et~al.}(2019)\citenamefont {Seo},
  \citenamefont {Kotov},\ and\ \citenamefont {Uchoa}}]{seo}%
  \BibitemOpen
  \bibfield  {author} {\bibinfo {author} {\bibfnamefont {Kangjun}\ \bibnamefont
  {Seo}}, \bibinfo {author} {\bibfnamefont {Valeri~N.}\ \bibnamefont {Kotov}},
  \ and\ \bibinfo {author} {\bibfnamefont {Bruno}\ \bibnamefont {Uchoa}},\
  }\bibfield  {title} {\enquote {\bibinfo {title} {Ferromagnetic mott state in
  twisted graphene bilayers at the magic angle},}\ }\href {\doibase
  10.1103/PhysRevLett.122.246402} {\bibfield  {journal} {\bibinfo  {journal}
  {Phys. Rev. Lett.}\ }\textbf {\bibinfo {volume} {122}},\ \bibinfo {pages}
  {246402} (\bibinfo {year} {2019})}\BibitemShut {NoStop}%
\bibitem [{\citenamefont {Kang}\ and\ \citenamefont {Vafek}(2019)}]{kang}%
  \BibitemOpen
  \bibfield  {author} {\bibinfo {author} {\bibfnamefont {Jian}\ \bibnamefont
  {Kang}}\ and\ \bibinfo {author} {\bibfnamefont {Oskar}\ \bibnamefont
  {Vafek}},\ }\bibfield  {title} {\enquote {\bibinfo {title} {Strong coupling
  phases of partially filled twisted bilayer graphene narrow bands},}\ }\href
  {\doibase 10.1103/PhysRevLett.122.246401} {\bibfield  {journal} {\bibinfo
  {journal} {Phys. Rev. Lett.}\ }\textbf {\bibinfo {volume} {122}},\ \bibinfo
  {pages} {246401} (\bibinfo {year} {2019})}\BibitemShut {NoStop}%
\bibitem [{\citenamefont {Lu}\ \emph {et~al.}(2019)\citenamefont {Lu},
  \citenamefont {Stepanov}, \citenamefont {Yang}, \citenamefont {Xie},
  \citenamefont {Aamir}, \citenamefont {Das}, \citenamefont {Urgell},
  \citenamefont {Watanabe}, \citenamefont {Taniguchi}, \citenamefont {Zhang},
  \citenamefont {Bachtold}, \citenamefont {MacDonald},\ and\ \citenamefont
  {Efetov}}]{efetov}%
  \BibitemOpen
  \bibfield  {author} {\bibinfo {author} {\bibfnamefont {Xiaobo}\ \bibnamefont
  {Lu}}, \bibinfo {author} {\bibfnamefont {Petr}\ \bibnamefont {Stepanov}},
  \bibinfo {author} {\bibfnamefont {Wei}\ \bibnamefont {Yang}}, \bibinfo
  {author} {\bibfnamefont {Ming}\ \bibnamefont {Xie}}, \bibinfo {author}
  {\bibfnamefont {Mohammed~Ali}\ \bibnamefont {Aamir}}, \bibinfo {author}
  {\bibfnamefont {Ipsita}\ \bibnamefont {Das}}, \bibinfo {author}
  {\bibfnamefont {Carles}\ \bibnamefont {Urgell}}, \bibinfo {author}
  {\bibfnamefont {Kenji}\ \bibnamefont {Watanabe}}, \bibinfo {author}
  {\bibfnamefont {Takashi}\ \bibnamefont {Taniguchi}}, \bibinfo {author}
  {\bibfnamefont {Guangyu}\ \bibnamefont {Zhang}}, \bibinfo {author}
  {\bibfnamefont {Adrian}\ \bibnamefont {Bachtold}}, \bibinfo {author}
  {\bibfnamefont {Allan~H.}\ \bibnamefont {MacDonald}}, \ and\ \bibinfo
  {author} {\bibfnamefont {Dmitri~K.}\ \bibnamefont {Efetov}},\ }\bibfield
  {title} {\enquote {\bibinfo {title} {Superconductors, orbital magnets and
  correlated states in magic-angle bilayer graphene},}\ }\href {\doibase
  10.1038/s41586-019-1695-0} {\bibfield  {journal} {\bibinfo  {journal}
  {Nature}\ }\textbf {\bibinfo {volume} {574}},\ \bibinfo {pages} {653--657}
  (\bibinfo {year} {2019})}\BibitemShut {NoStop}%
\bibitem [{\citenamefont {Padhi}\ \emph {et~al.}(2018)\citenamefont {Padhi},
  \citenamefont {Setty},\ and\ \citenamefont {Phillips}}]{philip1}%
  \BibitemOpen
  \bibfield  {author} {\bibinfo {author} {\bibfnamefont {Bikash}\ \bibnamefont
  {Padhi}}, \bibinfo {author} {\bibfnamefont {Chandan}\ \bibnamefont {Setty}},
  \ and\ \bibinfo {author} {\bibfnamefont {Philip~W.}\ \bibnamefont
  {Phillips}},\ }\bibfield  {title} {\enquote {\bibinfo {title} {Doped twisted
  bilayer graphene near magic angles: Proximity to wigner crystallization, not
  mott insulation},}\ }\href {\doibase 10.1021/acs.nanolett.8b02033} {\bibfield
   {journal} {\bibinfo  {journal} {Nano Letters}\ }\textbf {\bibinfo {volume}
  {18}},\ \bibinfo {pages} {6175–6180} (\bibinfo {year} {2018})}\BibitemShut
  {NoStop}%
\bibitem [{\citenamefont {Padhi}\ and\ \citenamefont
  {Phillips}(2019)}]{philip2}%
  \BibitemOpen
  \bibfield  {author} {\bibinfo {author} {\bibfnamefont {Bikash}\ \bibnamefont
  {Padhi}}\ and\ \bibinfo {author} {\bibfnamefont {Philip~W.}\ \bibnamefont
  {Phillips}},\ }\bibfield  {title} {\enquote {\bibinfo {title}
  {Pressure-induced metal-insulator transition in twisted bilayer graphene},}\
  }\href {\doibase 10.1103/physrevb.99.205141} {\bibfield  {journal} {\bibinfo
  {journal} {Physical Review B}\ }\textbf {\bibinfo {volume} {99}} (\bibinfo
  {year} {2019}),\ 10.1103/physrevb.99.205141}\BibitemShut {NoStop}%
\bibitem [{\citenamefont {Pizarro}\ \emph {et~al.}(2019)\citenamefont
  {Pizarro}, \citenamefont {Calder{\'{o}}n},\ and\ \citenamefont
  {Bascones}}]{pizarro}%
  \BibitemOpen
  \bibfield  {author} {\bibinfo {author} {\bibfnamefont {J~M}\ \bibnamefont
  {Pizarro}}, \bibinfo {author} {\bibfnamefont {M~J}\ \bibnamefont
  {Calder{\'{o}}n}}, \ and\ \bibinfo {author} {\bibfnamefont {E}~\bibnamefont
  {Bascones}},\ }\bibfield  {title} {\enquote {\bibinfo {title} {The nature of
  correlations in the insulating states of twisted bilayer graphene},}\ }\href
  {\doibase 10.1088/2399-6528/ab0fa9} {\bibfield  {journal} {\bibinfo
  {journal} {Journal of Physics Communications}\ }\textbf {\bibinfo {volume}
  {3}},\ \bibinfo {pages} {035024} (\bibinfo {year} {2019})}\BibitemShut
  {NoStop}%
\bibitem [{\citenamefont {Xie}\ and\ \citenamefont
  {MacDonald}(2020)}]{xiemacdonald}%
  \BibitemOpen
  \bibfield  {author} {\bibinfo {author} {\bibfnamefont {Ming}\ \bibnamefont
  {Xie}}\ and\ \bibinfo {author} {\bibfnamefont {A.~H.}\ \bibnamefont
  {MacDonald}},\ }\bibfield  {title} {\enquote {\bibinfo {title} {Nature of the
  correlated insulator states in twisted bilayer graphene},}\ }\href {\doibase
  10.1103/PhysRevLett.124.097601} {\bibfield  {journal} {\bibinfo  {journal}
  {Phys. Rev. Lett.}\ }\textbf {\bibinfo {volume} {124}},\ \bibinfo {pages}
  {097601} (\bibinfo {year} {2020})}\BibitemShut {NoStop}%
\bibitem [{\citenamefont {Repellin}\ \emph {et~al.}(2020)\citenamefont
  {Repellin}, \citenamefont {Dong}, \citenamefont {Zhang},\ and\ \citenamefont
  {Senthil}}]{repelin}%
  \BibitemOpen
  \bibfield  {author} {\bibinfo {author} {\bibfnamefont {C\'ecile}\
  \bibnamefont {Repellin}}, \bibinfo {author} {\bibfnamefont {Zhihuan}\
  \bibnamefont {Dong}}, \bibinfo {author} {\bibfnamefont {Ya-Hui}\ \bibnamefont
  {Zhang}}, \ and\ \bibinfo {author} {\bibfnamefont {T.}~\bibnamefont
  {Senthil}},\ }\bibfield  {title} {\enquote {\bibinfo {title} {Ferromagnetism
  in narrow bands of moir\'e superlattices},}\ }\href {\doibase
  10.1103/PhysRevLett.124.187601} {\bibfield  {journal} {\bibinfo  {journal}
  {Phys. Rev. Lett.}\ }\textbf {\bibinfo {volume} {124}},\ \bibinfo {pages}
  {187601} (\bibinfo {year} {2020})}\BibitemShut {NoStop}%
\bibitem [{\citenamefont {Koshino}\ \emph {et~al.}(2018)\citenamefont
  {Koshino}, \citenamefont {Yuan}, \citenamefont {Koretsune}, \citenamefont
  {Ochi}, \citenamefont {Kuroki},\ and\ \citenamefont {Fu}}]{koshino-fu}%
  \BibitemOpen
  \bibfield  {author} {\bibinfo {author} {\bibfnamefont {Mikito}\ \bibnamefont
  {Koshino}}, \bibinfo {author} {\bibfnamefont {Noah F.~Q.}\ \bibnamefont
  {Yuan}}, \bibinfo {author} {\bibfnamefont {Takashi}\ \bibnamefont
  {Koretsune}}, \bibinfo {author} {\bibfnamefont {Masayuki}\ \bibnamefont
  {Ochi}}, \bibinfo {author} {\bibfnamefont {Kazuhiko}\ \bibnamefont {Kuroki}},
  \ and\ \bibinfo {author} {\bibfnamefont {Liang}\ \bibnamefont {Fu}},\
  }\bibfield  {title} {\enquote {\bibinfo {title} {Maximally localized wannier
  orbitals and the extended hubbard model for twisted bilayer graphene},}\
  }\href {\doibase 10.1103/PhysRevX.8.031087} {\bibfield  {journal} {\bibinfo
  {journal} {Phys. Rev. X}\ }\textbf {\bibinfo {volume} {8}},\ \bibinfo {pages}
  {031087} (\bibinfo {year} {2018})}\BibitemShut {NoStop}%
\bibitem [{\citenamefont {Po}\ \emph {et~al.}(2018)\citenamefont {Po},
  \citenamefont {Zou}, \citenamefont {Vishwanath},\ and\ \citenamefont
  {Senthil}}]{zhouvish}%
  \BibitemOpen
  \bibfield  {author} {\bibinfo {author} {\bibfnamefont {Hoi~Chun}\
  \bibnamefont {Po}}, \bibinfo {author} {\bibfnamefont {Liujun}\ \bibnamefont
  {Zou}}, \bibinfo {author} {\bibfnamefont {Ashvin}\ \bibnamefont
  {Vishwanath}}, \ and\ \bibinfo {author} {\bibfnamefont {T.}~\bibnamefont
  {Senthil}},\ }\bibfield  {title} {\enquote {\bibinfo {title} {Origin of mott
  insulating behavior and superconductivity in twisted bilayer graphene},}\
  }\href {\doibase 10.1103/PhysRevX.8.031089} {\bibfield  {journal} {\bibinfo
  {journal} {Phys. Rev. X}\ }\textbf {\bibinfo {volume} {8}},\ \bibinfo {pages}
  {031089} (\bibinfo {year} {2018})}\BibitemShut {NoStop}%
\bibitem [{\citenamefont {Zou}\ \emph {et~al.}(2018)\citenamefont {Zou},
  \citenamefont {Po}, \citenamefont {Vishwanath},\ and\ \citenamefont
  {Senthil}}]{zou2}%
  \BibitemOpen
  \bibfield  {author} {\bibinfo {author} {\bibfnamefont {Liujun}\ \bibnamefont
  {Zou}}, \bibinfo {author} {\bibfnamefont {Hoi~Chun}\ \bibnamefont {Po}},
  \bibinfo {author} {\bibfnamefont {Ashvin}\ \bibnamefont {Vishwanath}}, \ and\
  \bibinfo {author} {\bibfnamefont {T.}~\bibnamefont {Senthil}},\ }\bibfield
  {title} {\enquote {\bibinfo {title} {Band structure of twisted bilayer
  graphene: Emergent symmetries, commensurate approximants, and wannier
  obstructions},}\ }\href {\doibase 10.1103/PhysRevB.98.085435} {\bibfield
  {journal} {\bibinfo  {journal} {Phys. Rev. B}\ }\textbf {\bibinfo {volume}
  {98}},\ \bibinfo {pages} {085435} (\bibinfo {year} {2018})}\BibitemShut
  {NoStop}%
\bibitem [{\citenamefont {{Wang}}\ and\ \citenamefont
  {{Vafek}}(2020)}]{vafek-wang}%
  \BibitemOpen
  \bibfield  {author} {\bibinfo {author} {\bibfnamefont {Xiaoyu}\ \bibnamefont
  {{Wang}}}\ and\ \bibinfo {author} {\bibfnamefont {Oskar}\ \bibnamefont
  {{Vafek}}},\ }\bibfield  {title} {\enquote {\bibinfo {title} {{A diagnosis of
  explicit symmetry breaking in the tight-binding constructions for
  symmetry-protected topological systems}},}\ }\href@noop {} {\bibfield
  {journal} {\bibinfo  {journal} {arXiv e-prints}\ } (\bibinfo {year}
  {2020})},\ \Eprint {http://arxiv.org/abs/2002.02057} {arXiv:2002.02057
  [cond-mat.str-el]} \BibitemShut {NoStop}%
\bibitem [{\citenamefont {Ewald}(1921)}]{ewald}%
  \BibitemOpen
  \bibfield  {author} {\bibinfo {author} {\bibfnamefont {P.~P.}\ \bibnamefont
  {Ewald}},\ }\bibfield  {title} {\enquote {\bibinfo {title} {Die berechnung
  optischer und elektrostatischer gitterpotentiale},}\ }\href {\doibase
  10.1002/andp.19213690304} {\bibfield  {journal} {\bibinfo  {journal} {Annalen
  der Physik}\ }\textbf {\bibinfo {volume} {369}},\ \bibinfo {pages} {253--287}
  (\bibinfo {year} {1921})}\BibitemShut {NoStop}%
\bibitem [{\citenamefont {Lee}\ and\ \citenamefont {Cai}(2009)}]{ewaldlee}%
  \BibitemOpen
  \bibfield  {author} {\bibinfo {author} {\bibfnamefont {H.}~\bibnamefont
  {Lee}}\ and\ \bibinfo {author} {\bibfnamefont {W.}~\bibnamefont {Cai}},\
  }\href {http://micro.stanford.edu/mediawiki/images/4/46/Ewald_notes.pdf}
  {\enquote {\bibinfo {title} {Ewald summation for coulomb interactions in a
  periodic supercell},}\ }\bibinfo {howpublished} {Lecture Notes, Stanford
  University} (\bibinfo {year} {2009})\BibitemShut {NoStop}%
\bibitem [{\citenamefont {Wu}\ \emph {et~al.}(2018)\citenamefont {Wu},
  \citenamefont {Lovorn}, \citenamefont {Tutuc},\ and\ \citenamefont
  {MacDonald}}]{PhysRevLett.121.026402}%
  \BibitemOpen
  \bibfield  {author} {\bibinfo {author} {\bibfnamefont {Fengcheng}\
  \bibnamefont {Wu}}, \bibinfo {author} {\bibfnamefont {Timothy}\ \bibnamefont
  {Lovorn}}, \bibinfo {author} {\bibfnamefont {Emanuel}\ \bibnamefont {Tutuc}},
  \ and\ \bibinfo {author} {\bibfnamefont {A.~H.}\ \bibnamefont {MacDonald}},\
  }\bibfield  {title} {\enquote {\bibinfo {title} {Hubbard model physics in
  transition metal dichalcogenide moir\'e bands},}\ }\href {\doibase
  10.1103/PhysRevLett.121.026402} {\bibfield  {journal} {\bibinfo  {journal}
  {Phys. Rev. Lett.}\ }\textbf {\bibinfo {volume} {121}},\ \bibinfo {pages}
  {026402} (\bibinfo {year} {2018})}\BibitemShut {NoStop}%
\bibitem [{\citenamefont {Wu}\ \emph {et~al.}(2019)\citenamefont {Wu},
  \citenamefont {Lovorn}, \citenamefont {Tutuc}, \citenamefont {Martin},\ and\
  \citenamefont {MacDonald}}]{PhysRevLett.122.086402}%
  \BibitemOpen
  \bibfield  {author} {\bibinfo {author} {\bibfnamefont {Fengcheng}\
  \bibnamefont {Wu}}, \bibinfo {author} {\bibfnamefont {Timothy}\ \bibnamefont
  {Lovorn}}, \bibinfo {author} {\bibfnamefont {Emanuel}\ \bibnamefont {Tutuc}},
  \bibinfo {author} {\bibfnamefont {Ivar}\ \bibnamefont {Martin}}, \ and\
  \bibinfo {author} {\bibfnamefont {A.~H.}\ \bibnamefont {MacDonald}},\
  }\bibfield  {title} {\enquote {\bibinfo {title} {Topological insulators in
  twisted transition metal dichalcogenide homobilayers},}\ }\href {\doibase
  10.1103/PhysRevLett.122.086402} {\bibfield  {journal} {\bibinfo  {journal}
  {Phys. Rev. Lett.}\ }\textbf {\bibinfo {volume} {122}},\ \bibinfo {pages}
  {086402} (\bibinfo {year} {2019})}\BibitemShut {NoStop}%
\bibitem [{\citenamefont {Tang}\ \emph {et~al.}(2020)\citenamefont {Tang},
  \citenamefont {Li}, \citenamefont {Li}, \citenamefont {Xu}, \citenamefont
  {Liu}, \citenamefont {Barmak}, \citenamefont {Watanabe}, \citenamefont
  {Taniguchi}, \citenamefont {MacDonald}, \citenamefont {Shan},\ and\
  \citenamefont {Mak}}]{2019arXiv191008673T}%
  \BibitemOpen
  \bibfield  {author} {\bibinfo {author} {\bibfnamefont {Yanhao}\ \bibnamefont
  {Tang}}, \bibinfo {author} {\bibfnamefont {Lizhong}\ \bibnamefont {Li}},
  \bibinfo {author} {\bibfnamefont {Tingxin}\ \bibnamefont {Li}}, \bibinfo
  {author} {\bibfnamefont {Yang}\ \bibnamefont {Xu}}, \bibinfo {author}
  {\bibfnamefont {Song}\ \bibnamefont {Liu}}, \bibinfo {author} {\bibfnamefont
  {Katayun}\ \bibnamefont {Barmak}}, \bibinfo {author} {\bibfnamefont {Kenji}\
  \bibnamefont {Watanabe}}, \bibinfo {author} {\bibfnamefont {Takashi}\
  \bibnamefont {Taniguchi}}, \bibinfo {author} {\bibfnamefont {Allan~H.}\
  \bibnamefont {MacDonald}}, \bibinfo {author} {\bibfnamefont {Jie}\
  \bibnamefont {Shan}}, \ and\ \bibinfo {author} {\bibfnamefont {Kin~Fai}\
  \bibnamefont {Mak}},\ }\bibfield  {title} {\enquote {\bibinfo {title}
  {Simulation of hubbard model physics in wse2/ws2 moir{\'e} superlattices},}\
  }\href {\doibase 10.1038/s41586-020-2085-3} {\bibfield  {journal} {\bibinfo
  {journal} {Nature}\ }\textbf {\bibinfo {volume} {579}},\ \bibinfo {pages}
  {353--358} (\bibinfo {year} {2020})}\BibitemShut {NoStop}%
\bibitem [{\citenamefont {{Regan}}\ \emph {et~al.}(2020)\citenamefont
  {{Regan}}, \citenamefont {{Wang}}, \citenamefont {{Jin}}, \citenamefont
  {{Bakti Utama}}, \citenamefont {{Gao}}, \citenamefont {{Wei}}, \citenamefont
  {{Zhao}}, \citenamefont {{Zhao}}, \citenamefont {{Zhang}}, \citenamefont
  {{Yumigeta}}, \citenamefont {{Blei}}, \citenamefont {{Carlstr{\"o}m}},
  \citenamefont {{Watanabe}}, \citenamefont {{Taniguchi}}, \citenamefont
  {{Tongay}}, \citenamefont {{Crommie}}, \citenamefont {{Zettl}},\ and\
  \citenamefont {{Wang}}}]{2020Natur.579..359R}%
  \BibitemOpen
  \bibfield  {author} {\bibinfo {author} {\bibfnamefont {Emma~C.}\ \bibnamefont
  {{Regan}}}, \bibinfo {author} {\bibfnamefont {Danqing}\ \bibnamefont
  {{Wang}}}, \bibinfo {author} {\bibfnamefont {Chenhao}\ \bibnamefont {{Jin}}},
  \bibinfo {author} {\bibfnamefont {M.~Iqbal}\ \bibnamefont {{Bakti Utama}}},
  \bibinfo {author} {\bibfnamefont {Beini}\ \bibnamefont {{Gao}}}, \bibinfo
  {author} {\bibfnamefont {Xin}\ \bibnamefont {{Wei}}}, \bibinfo {author}
  {\bibfnamefont {Sihan}\ \bibnamefont {{Zhao}}}, \bibinfo {author}
  {\bibfnamefont {Wenyu}\ \bibnamefont {{Zhao}}}, \bibinfo {author}
  {\bibfnamefont {Zuocheng}\ \bibnamefont {{Zhang}}}, \bibinfo {author}
  {\bibfnamefont {Kentaro}\ \bibnamefont {{Yumigeta}}}, \bibinfo {author}
  {\bibfnamefont {Mark}\ \bibnamefont {{Blei}}}, \bibinfo {author}
  {\bibfnamefont {Johan~D.}\ \bibnamefont {{Carlstr{\"o}m}}}, \bibinfo {author}
  {\bibfnamefont {Kenji}\ \bibnamefont {{Watanabe}}}, \bibinfo {author}
  {\bibfnamefont {Takashi}\ \bibnamefont {{Taniguchi}}}, \bibinfo {author}
  {\bibfnamefont {Sefaattin}\ \bibnamefont {{Tongay}}}, \bibinfo {author}
  {\bibfnamefont {Michael}\ \bibnamefont {{Crommie}}}, \bibinfo {author}
  {\bibfnamefont {Alex}\ \bibnamefont {{Zettl}}}, \ and\ \bibinfo {author}
  {\bibfnamefont {Feng}\ \bibnamefont {{Wang}}},\ }\bibfield  {title} {\enquote
  {\bibinfo {title} {{Mott and generalized Wigner crystal states in
  WSe$_{2}$/WS$_{2}$ moir{\'e} superlattices}},}\ }\href {\doibase
  10.1038/s41586-020-2092-4} {\bibfield  {journal} {\bibinfo  {journal} {\nat}\
  }\textbf {\bibinfo {volume} {579}},\ \bibinfo {pages} {359--363} (\bibinfo
  {year} {2020})}\BibitemShut {NoStop}%
\bibitem [{\citenamefont {{Zhang}}\ \emph {et~al.}(2019)\citenamefont
  {{Zhang}}, \citenamefont {{Yuan}},\ and\ \citenamefont {{Fu}}}]{fu-tmd}%
  \BibitemOpen
  \bibfield  {author} {\bibinfo {author} {\bibfnamefont {Yang}\ \bibnamefont
  {{Zhang}}}, \bibinfo {author} {\bibfnamefont {Noah F.~Q.}\ \bibnamefont
  {{Yuan}}}, \ and\ \bibinfo {author} {\bibfnamefont {Liang}\ \bibnamefont
  {{Fu}}},\ }\bibfield  {title} {\enquote {\bibinfo {title} {{Moir{\'e} quantum
  chemistry: charge transfer in transition metal dichalcogenide
  superlattices}},}\ }\href@noop {} {\bibfield  {journal} {\bibinfo  {journal}
  {arXiv e-prints}\ } (\bibinfo {year} {2019})},\ \Eprint
  {http://arxiv.org/abs/1910.14061} {arXiv:1910.14061 [cond-mat.str-el]}
  \BibitemShut {NoStop}%
\bibitem [{\citenamefont {Wang}\ \emph {et~al.}(2020)\citenamefont {Wang},
  \citenamefont {Shih}, \citenamefont {Ghiotto}, \citenamefont {Xian},
  \citenamefont {Rhodes}, \citenamefont {Tan}, \citenamefont {Claassen},
  \citenamefont {Kennes}, \citenamefont {Bai}, \citenamefont {Kim},
  \citenamefont {Watanabe}, \citenamefont {Taniguchi}, \citenamefont {Zhu},
  \citenamefont {Hone}, \citenamefont {Rubio}, \citenamefont {Pasupathy},\ and\
  \citenamefont {Dean}}]{cory-wse2}%
  \BibitemOpen
  \bibfield  {author} {\bibinfo {author} {\bibfnamefont {Lei}\ \bibnamefont
  {Wang}}, \bibinfo {author} {\bibfnamefont {En-Min}\ \bibnamefont {Shih}},
  \bibinfo {author} {\bibfnamefont {Augusto}\ \bibnamefont {Ghiotto}}, \bibinfo
  {author} {\bibfnamefont {Lede}\ \bibnamefont {Xian}}, \bibinfo {author}
  {\bibfnamefont {Daniel~A.}\ \bibnamefont {Rhodes}}, \bibinfo {author}
  {\bibfnamefont {Cheng}\ \bibnamefont {Tan}}, \bibinfo {author} {\bibfnamefont
  {Martin}\ \bibnamefont {Claassen}}, \bibinfo {author} {\bibfnamefont
  {Dante~M.}\ \bibnamefont {Kennes}}, \bibinfo {author} {\bibfnamefont
  {Yusong}\ \bibnamefont {Bai}}, \bibinfo {author} {\bibfnamefont {Bumho}\
  \bibnamefont {Kim}}, \bibinfo {author} {\bibfnamefont {Kenji}\ \bibnamefont
  {Watanabe}}, \bibinfo {author} {\bibfnamefont {Takashi}\ \bibnamefont
  {Taniguchi}}, \bibinfo {author} {\bibfnamefont {Xiaoyang}\ \bibnamefont
  {Zhu}}, \bibinfo {author} {\bibfnamefont {James}\ \bibnamefont {Hone}},
  \bibinfo {author} {\bibfnamefont {Angel}\ \bibnamefont {Rubio}}, \bibinfo
  {author} {\bibfnamefont {Abhay~N.}\ \bibnamefont {Pasupathy}}, \ and\
  \bibinfo {author} {\bibfnamefont {Cory~R.}\ \bibnamefont {Dean}},\ }\bibfield
   {title} {\enquote {\bibinfo {title} {Correlated electronic phases in twisted
  bilayer transition metal dichalcogenides},}\ }\href
  {https://doi.org/10.1038/s41563-020-0708-6} {\bibfield  {journal} {\bibinfo
  {journal} {Nature Materials}\ } (\bibinfo {year} {2020})}\BibitemShut
  {NoStop}%
\end{thebibliography}%

\end{document}